\documentstyle[12pt]{article}

\newcommand{\nc}{\newcommand}
\nc{\beq}{\begin{equation}}
\nc{\eeq}{\end{equation}}
\nc{\beqa}{\begin{eqnarray}}
\nc{\eeqa}{\end{eqnarray}}

\def\O {{\cal O}}
\def\ts{\thinspace}
\def\fpi{F}

\input epsf
\newwrite\ffile\global\newcount\figno \global\figno=1

\def\writedef#1{}
\def\figin{\epsfcheck\figin}\def\figins{\epsfcheck\figins}
\def\epsfcheck{\ifx\epsfbox\UnDeFiNeD
\message{(NO epsf.tex, FIGURES WILL BE IGNORED)}
\gdef\figin##1{\vskip2in}\gdef\figins##1{\hskip.5in}
\else\message{(FIGURES WILL BE INCLUDED)}%
\gdef\figin##1{##1}\gdef\figins##1{##1}\fi}
\def\figinsert{}
\def\ifig#1#2#3{\xdef#1{fig.~\the\figno}
\writedef{#1\leftbracket fig.\noexpand~\the\figno}%
\figinsert\figin{\centerline{#3}}\medskip\centerline{\vbox{\baselineskip12pt
\advance\hsize by -1truein\center\footnotesize{  Fig.~\the\figno.} #2}}
\bigskip\endinsert\global\advance\figno by1}
\def\endinsert{}

\begin{document}



\title{\large{\bf Particle Multiplicities and Thermalization
in High Energy Collisions }}

\author{
James Hormuzdiar$^1$\thanks{hormuzj@muller.physics.mcgill.ca},
Stephen D.H.~Hsu$^2$\thanks{hsu@duende.uoregon.edu},
Gregory Mahlon$^1$\thanks{mahlon@physics.mcgill.ca}\\
\\
$^1$Department of Physics, \\
McGill University, 3600 University Street, \\
Montr\' eal, Qu\' ebec, H3A 2T8, Canada \\
\\
$^2$Department of Physics, \\
University of Oregon, Eugene OR 97403-5203 \\ \\   }


\maketitle

\begin{picture}(0,0)(0,0)
\put(350,380){McGill/00-04}
\put(350,365){OITS-686}
\end{picture}
\vspace{-24pt}

\begin{abstract}
We investigate the conditions under which particle multiplicities
in high energy collisions are Boltzmann distributed, as is the
case for hadron production in $e^+e^-$, $pp$, $p \bar p$ and heavy
ion collisions. We show that the {\it apparent} temperature
governing this distribution does not necessarily imply equilibrium
(thermal or chemical) in the usual sense, as we explain. We
discuss an explicit example using tree level amplitudes for $N$
photon production in which a Boltzmann-like distribution is
obtained without any equilibration. We argue that the {\it
failure} of statistical techniques based on free particle
ensembles may provide a signal for collective phenomena (such as
large shifts in masses and widths of resonances) related to the
QCD phase transition.

\end{abstract}


\newpage

\section{Introduction}

The use of thermal or statistical models to describe multiparticle
production has a long history \cite{old}. Recent
studies~[\ref{BH1}--\ref{BH8}] observe that the multiplicities of
hadron production in a variety of contexts ($e^+e^-$, $pp$, $p\bar
p$ and heavy ion collisions) are extremely well described by
models involving thermal distributions of free hadrons. The only
free parameters in their analysis are the temperature and volume
of the model thermal system, and a parameter reflecting the level
of equilibration of strange particles (in the heavy ion case the
baryon chemical potential is an additional parameter). One might
conclude from their results that thermal and chemical equilibrium
among hadrons is already reached in individual jets produced in
high-energy scattering.

However, this conclusion is somewhat unjustified, given that {\it
any} mechanism for producing hadrons which evenly populates the
free particle phase space will mimic a microcanonical ensemble,
and therefore yield {\it apparently} thermal
results\footnote{After completion of this work we learned that a
similar conclusion was reached by C. N. Yang et al. in
\cite{Yang}. They refer to the apparent temperature as a {\it
partition temperature} and stress that no thermal equilibrium is
implied. We thank Professor K.S. Lee of Chonnam National
University in Korea for making us aware of this earlier work.}. It
is important to remark that this type of {\it apparent}
thermalization will always yield Boltzmann weights which are
functions of the free particle energy, and hence describe a
non-interacting ensemble. True QCD thermalization, of the type
associated with the quark-gluon plasma (QGP), and that
experimentalists hope to observe at RHIC, involves large
collective effects, and hence is probably poorly modelled by a
non-interacting hadron gas.

Here we address the problem of deducing whether a process which
leads to multiparticle production is thermal. We make the important
distinction between phase space dominated phenomena, which lead to
ensembles governed by free particle Boltzmann weights, and interacting
thermal ensembles, in which collective effects can be important. We argue that
a process which generates data that can be fit using free particle ensembles
is merely good at populating phase space in a uniform way
-- it has not necessarily
produced an interacting thermal region. In QCD, a region of this type
with temperature
of order 100~MeV or more should exhibit strong collective phenomena, such
as hadron mass shifts, that preclude description in terms of a free particle
ensemble. Hence, we argue that the {\it failure} of statistical techniques
based on free particle ensembles should
be regarded as a signal for the onset of true equilibration at heavy
ion colliders such as RHIC.

This paper is organized as follows. In section 2 we review the
relation between microcanonical and canonical ensembles in
statistical mechanics. We argue that multiparticle production in
many cases is equivalent to a method of populating a modified
microcanonical ensemble. Such an ensemble will produce ``thermal''
behavior even if there is no subsequent interaction of particles
once produced. In section 3 we apply our results to multiparticle
production and examine a specific example in a toy model involving
tree-level photon production.  In this toy model there is clearly
no real thermal equilibrium -- once produced, the photons do not
interact. However, a quasi-Boltzmann distribution results in the
limit where a large number of photons is produced.  In section 4
we discuss the implications of our conclusions for heavy ion
collisions.


\section{Microcanonical vs Canonical Ensembles}

Here we give a brief review of the
relationship
between the
microcanonical and canonical ensembles (MCE and CE, respectively)
in statistical mechanics.
Recall that the MCE sums only over states with some fixed total
energy, while the CE sums over all states with Boltzmann weight
$e^{- \beta E}$.
The main result is rather familiar: under certain general assumptions,
quantities computed in the MCE differ from those computed in
the CE by an amount which vanishes as the size of the system $N$
is taken to infinity.
The importance of this result is as follows: the cross section for
particle production in a high energy collision can be written as the
expectation of the matrix element squared in the MCE corresponding
to the {\it free} theory.  In the limit
of large $N$, one can therefore rewrite the usual phase
space integral appearing in a cross section in terms of the average of the
matrix element squared in a CE, which is controlled by the
Boltzmann factor. This naturally leads to a certain
``thermal'' behavior which we discuss below.

In the MCE the total energy is fixed and the
probability density is constant over all of phase space. The
computation of the entropy $S(E)$ is as follows:
\beq
\label{MCE}
\Gamma (E) = \sum_s \delta( E_s - E ),
\eeq
\beq
S(E) = \ln \Gamma (E).
\eeq
In Eq.~(\ref{MCE}), $E_s$ is the energy of the state $s$.
It is often more convenient to replace the delta function in (\ref{MCE})
with the factor
$\Delta (E) \equiv \exp ( - \beta (E_s - E))$, yielding a new quantity,
the CE:
\beq
\label{CE}
\bar{\Gamma} (E) = \sum_s e^{ - \beta (E_s - E)}.
\eeq
In the canonical ensemble there is no restriction on the energies of
the states $s$. However, they appear with Boltzmann weight
$\exp (- \beta E_s)$. The new quantity introduced, $\beta$, describes
the temperature of the system, which is fine-tuned to ensure that
the average energy is $E$. To see this, rewrite $\bar{\Gamma} (E)$ as
follows:
\beqa
\bar{\Gamma} (E) &=& \int dE'~ \sum_s e^{ - \beta (E' - E)} \delta( E_s -E' )
\nonumber \\
&=& \int dE'~ \Gamma(E') ~e^{- \beta (E' -E)} \nonumber \\
&=& \int dE'~ e^{S(E') - \beta (E' -E)} \label{CE1}. \eeqa Now
evaluate (\ref{CE1}) in the saddle-point approximation. (This is
also known as the Darwin-Fowler method \cite{Yang}.) Let
\beq
\label{temp} \beta ~=~ {\partial S \over \partial E'} \Bigg|_E ~~
, \eeq so \beqa \bar{\Gamma} &=& e^{S(E)} ~\int dE'~
e^{\frac{1}{2} S''(E) (E' -E)^2 + \cdots}
 \nonumber \\
 &=& \sqrt{ 2 \pi \over - S''(E)}~ e^{S(E)} ~+~ \cdots~~~.
\label{sp}
\eeqa
Here we have assumed that $S'(E) > 0$ (positive temperature) and
$S'' (E) < 0$ (positive specific heat).
It is easy to see that the difference between the CE entropy
$\bar{S}(E) = \ln \bar{\Gamma} (E)$ and the MCE entropy $S(E)$ is
of order $(\ln N)/N$. The terms represented by the ellipsis
in (\ref{sp}) lead to even smaller corrections, and will be neglected.

Now consider a generic operator $\O$. The average in the CE is given by
\beqa
\langle \O \rangle_{C} &=& {1 \over \bar{\Gamma}(E)} \sum_s
e^{- \beta (E_s -E)} ~\O_s \nonumber \\
&=& e^{-S(E)} \sqrt{ - S''(E) \over 2 \pi }~ \int dE'~
e^{S(E') - \beta (E' -E)}
\ts\ts\langle \O \rangle_{M} (E') \label{OCE},
\eeqa
where $\langle \O \rangle_{M}$ is the average taken in the microcanonical
ensemble. Its logarithm can be expanded as follows
\beq
\ln [\langle \O \rangle_{M} (E')] = \ln [\langle \O \rangle_{M} (E)] ~+~
(E'-E)  \ln'[ \langle \O \rangle_{M} (E)]  ~+~
 \frac{1}{2} (E'-E)^2 \ln'' [\langle \O \rangle_{M} (E)]  ~+~ \cdots .
\label{lnO} \eeq The integral in (\ref{OCE}) can again be
performed in the saddle-point approximation. Note that if the
operator $\O$ is of order $N$ (as in the case of a particle
multiplicity), the coefficients appearing in the expansion
(\ref{lnO}) are of order $\ln N$. They lead to small shifts (of
order $(\ln N)/N$) in the saddle-point value of the temperature
$\beta$ and the overall prefactor. Thus the canonical and
microcanonical averages of particle multiplicities converge in the
limit of large $N$.

In the next section we apply these results to the computation of
cross sections for particle production in high energy collisions.


\section{Multiplicity Results}

The cross section for the process $A+B \rightarrow n$ particles is
\beq
\label{sig}
\sigma = {1 \over 2 E_A 2 E_B |v_A - v_B|}
\int \prod_f {d^3 p_f \over (2 \pi)^3} {1 \over 2 E_f} ~
|{\cal M}|^2 ~
(2 \pi)^4 \delta^{(4)}\biggl(p_A+p_B-\sum_f p_f\biggr).
\label {cross_section}
\eeq
First, let us assume that
the function\footnote{In relativistic field theory, we
usually adopt the normalization convention
$\langle p \vert p' \rangle = 2 E_p ~(2 \pi)^3 \delta^3 (p - p')$,
which leads to the factor of $1/2 E_p$ in the phase space
density.   On the other hand, in statistical physics, we
usually adopt an energy-independent normalization.
Although the final result for the cross section is independent
of our normalization convention, what we mean by
``phase space dominated'' depends on what we choose to be the
unit of phase space.
In this paper we will always be referring to the statistical
mechanical unit of phase space, {\it i.e.}\ $d^3 p$.}
$ |{\cal M}|^2  / \prod 2 E_f $
in the dominant region of phase space is slowly
varying (we will relax this assumption shortly).
Then, we can use the following approximation
\beq
\label{sig1}
\sigma =
{1 \over 2 E_A 2 E_B |v_A - v_B|}~
{ |\bar{\cal M}|^2   \over \prod_f 2 \bar{E}_f }
\int \prod_f {d^3 p_f \over (2 \pi)^3} ~
(2 \pi)^4 \delta^{(4)}\biggl(p_A+p_B-\sum_f p_f\biggr),
\eeq
where $\bar{\cal M}, \bar{E}$ are averaged quantities.
As noted, the integral in (\ref{sig1}) is just the
microcanonical ensemble for $N$ free particles, and hence leads to
``thermal'' properties of the particle distributions and multiplicities.
Considering the more general case where the number of particles is not
fixed, we simply sum over all cross sections,
\beq
\sigma = \sum_n \sigma_{AB \rightarrow n},
\eeq
to obtain a MCE without fixed particle number. In the usual thermodynamic
limit this sum is dominated by some particular value of $n$, so it is
equivalent to consider the earlier case with $n=N$.

We now
treat the matrix element more carefully,
by retaining it in the
phase space integral. The resulting integral can still
be turned into a canonical ensemble using the result of the previous
section, provided the modified ``entropy'' ({\it i.e.},
the logarithm of the phase space integral including the matrix element)
continues to grow with total energy $E$ (positive temperature),
and the second derivative of this entropy
with respect to $E$ is negative
(positive specific heat\footnote{These requirements are satisfied
in the toy model we consider
below, where the modified entropy behaves as
$S(E) \sim N \ln \ln (E/m)$, for $N$ ``photons'' with total energy $E$.}).
The energy-momentum delta function can then be replaced by a
Boltzmann weight
$\exp[(p_A{+}p_B{-}\sum p_f) \cdot \beta]$,
which in the center of mass frame reduces to
$\exp[\beta ( E {-} \sum E_f)]$.
This yields the
following result for the differential cross section:
\beq
\label{dcs}
{d^3 \sigma \over d p^3_i} \propto
{1 \over E_i} \ts e^{-E_i/T}
\int \prod_{f \neq i} d^3 p_f \ts {1 \over E_f} ~
|{\cal M}|^2 ~
\exp \left( {-\sum_{f \neq i} E_f/T} \right).
\eeq
Note the natural appearance of the Boltzmann factor in (\ref{dcs}).

To proceed further, we need a specific model for the behavior of the
matrix element. There are very few cases in which the matrix element
for $N$-particle production is explicitly known.
One such case is the QED process
\beq
f \bar{f} \rightarrow 1 \enspace\hbox{spin up photon}
+ (n-1)\enspace\hbox{spin down photons},
\label{QEDproc}
\eeq
where the (massless) fermions have opposite spin.
The matrix element squared for~(\ref{QEDproc}) is~\cite{QED}
\beq
|{\cal M}(P_{\uparrow},Q_{\downarrow};
1_{\uparrow},2_{\downarrow},\cdots,n_{\downarrow})|^2
= (2e^2)^n \ts\ts
{
{(2Q\cdot p_1)^2}
\over
{(2P\cdot Q)}
} \ts
\prod^n _{i=2}
{
{ (2P\cdot Q) }
\over
{ (2P\cdot p_i)(2p_i\cdot Q) }
},
\eeq
where $P$, $Q$ and $p_i$ are the momenta of the two incoming fermions
and $n$ outgoing gauge particles respectively.
In the $(P, Q)$ center of momentum frame this may be written as
\beq
|{\cal M}(P_{\uparrow},Q_{\downarrow};
1_{\uparrow},2_{\downarrow},\cdots,n_{\downarrow})|^2
= {\cal K}  E_1^2 (1+\cos\theta_1)^2
\prod^n _{i=2}
{
{1}\over{E^2_i \sin ^2 \theta_i}
},
\eeq
where $E_i, \theta_i$ are the energy and production angle of the
$i$th photon, and we have lumped all of the remaining constant
factors into ${\cal K}$.
Using this to solve for the differential cross section yields
($i \neq 1$)
\beq
d ^3 \sigma \propto
{
{1}
\over
{E^3_i \sin ^2 \theta}
}\ts
e^{-E_i/T}\ts d^3 p_i.
\eeq
The resulting number density of (spin down) photons is then
\beq
\label{n1}
n \propto \int {d^3 p \over (2 \pi)^3} ~ {1 \over E^3 ~ \sin ^2 \theta} ~
e^{-E/T}.
\eeq
In this particular example we have a problem, because the photons
are massless and there is an infrared catastrophe due to arbitrarily
soft photons.  Of course, the number density of {\it observable}\
photons -- those with some minimum energy and angular separation from
the initial $f\bar{f}$ pair -- is finite.
For our purposes,
we can always eliminate this problem by introducing
a photon mass by hand in our toy model.
In fact, we can introduce several
species of ``photons'' with masses $m_i$. Then, the abundance
of each species is given by\footnote{In writing Eq.~(\protect\ref{toy})
we have chosen to ignore the angular dependence.  Had we chosen
to retain it, the angular integration would produce the factor
$\ln\bigl[(1+\protect\sqrt{1-4m_i^2/s})
 /(1-\protect\sqrt{1-4m_i^2/s})\bigr]$,
where $\protect\sqrt{s}$ is the center of mass energy of the collision.}
\beq
\label{n2}
n_i ~\propto~  \int {d^3 p \over (2 \pi)^3} ~
{ 1 \over (p^2 + m_i^2)^{3/2} } ~
e^{- \sqrt{p^2 + m_i^2} / T}.
\label{toy}
\eeq
The integral in (\ref{n2}) differs from the one appearing in a pure
Boltzmann distribution due to the factor of
$(p^2 + m_i^2)^{-3/2}$. Without this extra factor, the
integral simply reduces to $m_i^2 T K_2( m_i/T)$, where
$K_2(x)$ denotes the modified Bessel function of order 2.

\begin{figure}[t!]
\includegraphics{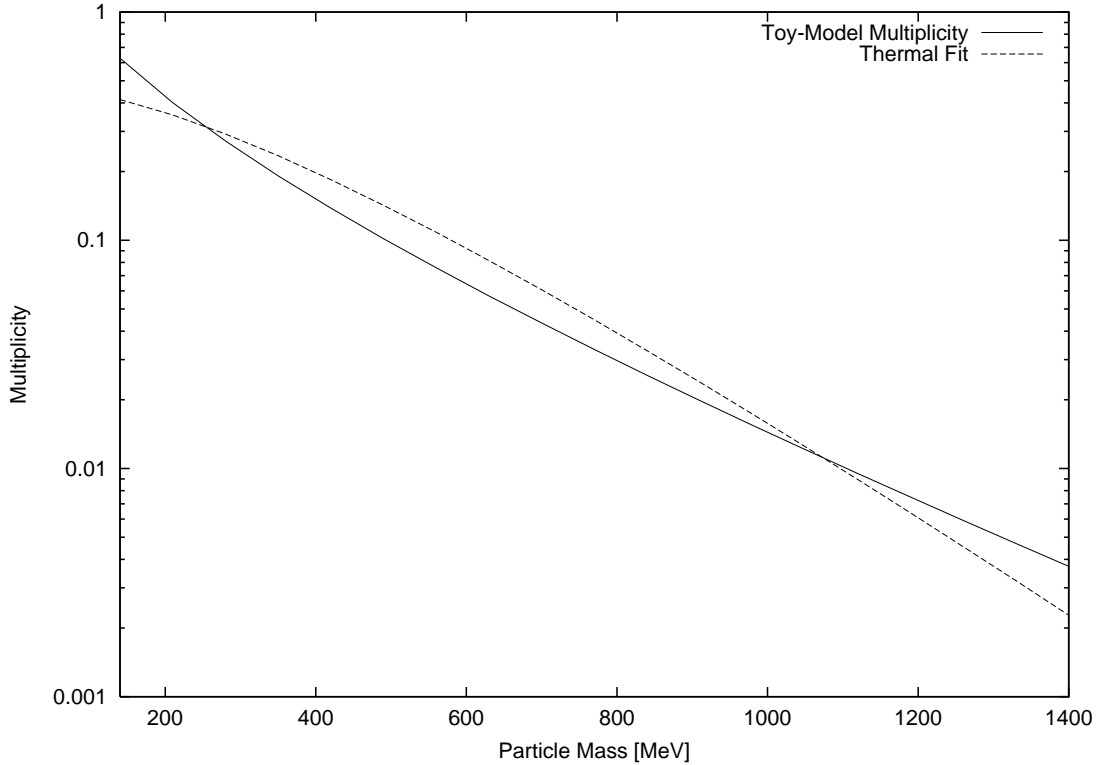}
\vspace{4.0in}
\caption{Multiplicity vs. particle species mass of the toy-model, and
thermal fit.}
\label{thermfit}
\end{figure}

In a sense, the additional
factor makes only a small difference relative to the exponential:
when taken into the exponent it is of order
$\ln(\beta E)$, versus $\beta E$ for the Boltzmann factor. However, actual
ratios of particle abundances $n_i / n_j$ will differ from thermal
ratios. In Fig.~\ref{thermfit} we show the result of the
multiplicities from (\ref{n2}) and a thermal best fit.
While we didn't use any additional parameters, such as individual
chemical potentials, the eventual quality of the fit would probably
not be as good as what is observed in $e^+e^-$, $pp$, $p\bar p$ and heavy ion
collisions~[\ref{BH1}--\ref{BH8}].
In other words, the hadronization process
probably populates free particle phase space somewhat more evenly than our
toy model. However, our toy model does demonstrate
that Boltzmann-like distributions are not necessarily indicative of
real thermal (chemical or kinetic) equilibrium.


\section{Discussion}

In the previous sections we argued that multiparticle production can
readily lead to thermal behavior if the the process in question is
phase space dominated. Because phase space is determined by
free particle kinematics,
the results correspond to an ensemble of non-interacting
particles. In other words, because the arguments of the energy
delta function in the cross section Eq.~(\ref{sig}) are simply
free particle energies, the corresponding Hamiltonian appearing
in the ensemble in Eqs.~(\ref{MCE}) and~(\ref{CE}) is the free
Hamiltonian, with no interactions. Our toy model of photons suggests
that this result is rather generic in any process where a large number
of particles is produced.

If the ensemble is dominated by a particular species of particle
({\it i.e.}\ the lightest particle),
the apparent ``temperature'' will be related to the mass $m$
of that species. This is
because phase space is maximized by producing as many particles as possible,
each with a kinetic energy of order $m$.
Suspiciously, the typical temperatures produced
by the excellent thermal fits of $e^+e^-$, $pp$, $p\bar p$ and heavy ion
collisions~[\ref{BH1}--\ref{BH8}]
are all of order the pion mass.

Of course, an ensemble of non-interacting hadrons is not very interesting.
It gives us {\it no} information about the actual QCD phase diagram.
In real QCD the energy of a state consisting of many
particles is modified due to interactions: it is not simply the
sum of the free particle energies.  At high density or temperature
interaction effects are
large and lead to large deviations from free particle results. Were this
not so one could never see collective phenomena such as chiral symmetry
restoration or a deconfinement phase transition.
In an ensemble of interacting hadrons we already expect
significant
collective effects at temperatures $T \sim m_\pi$,
such as a decrease in the value of the quark condensate
$\langle \bar{q} q \rangle_T$, and shifts in the various hadron masses.

We expect these effects to lead to the failure of statistical
techniques based on free particle ensembles,  which predict that
the multiplicities should fall roughly exponentially
(as $e^{-m/T}$, see Fig.~\ref{thermfit}).
However, if the masses and widths are shifted from their
vacuum values at the instant of chemical freeze-out, as would
be expected at the $\sim 170$ MeV temperatures obtained in
the fits, then a plot of multiplicity versus {\it vacuum}\ mass
will deviate from the thermal prediction:  {\it i.e.}\ it
will have the form\footnote{We have simplified our discussion here
in two ways.  First, we have ignored the possiblity of introducing
chemical potentials related to conserved quantities.  Since
the number of data points to be fit is much larger than the handful
of these potentials which may justifiably be introduced,  it is
highly unlikely that the effects of all of the mass shifts could be
reproduced in this manner.
Second, many of the hadronic states decay rapidly
and so affect the relative populations of the observed particles.
Inclusion of these effects adds no degrees of freedom to the
fits.  It would be amazing if the shifts in masses and widths
should conspire to reproduce the vacuum results.
}
$e^{-m(T)/T}$.
On the other hand, it is also possible that the system will
remain in equilibrium long enough for the masses and widths
to return to their vacuum values.  If so, then a thermal
fit will perform well, but the resulting
temperature would not be near the chiral phase transition.

Knowledge of these mass shifts would be necessary for any
detailed predictions.  Although there have been some
promising results~\cite{lattice}, it has proven difficult to
extract this information from lattice data.
A perturbative estimate of the
decrease in the quark condensate was
derived in Ref.~\cite{temp}.
In the chiral limit ($m_u=m_d=0$), an analytic result is possible.
It reads
\beq
\langle \bar{q} q \rangle_T ~=~ \langle \bar{q} q \rangle_{T=0}
\Biggl[ 1 - { 3 T^2 \over 24 \fpi^2} - \frac{3}{8}
\Biggl( {T^2 \over 12 \fpi^2} \Biggr)^2 + {\cal O}(T^6) \Biggr],
\eeq
where $\fpi \approx 93$ MeV is the (zero temperature) pion decay constant.
Using
real world quark masses changes this
result only slightly.
The corresponding estimate for the temperature of the chiral phase
transition ({\it i.e.}\ the point at which the condensate is
essentially zero) is 170~MeV in the chiral limit, and 190~MeV in the
real world,
consistent with lattice data.

We expect the thermal pion mass to obey the finite temperature
version of the Dashen formula
\beq
\label{DT}
m^2_\pi (T) ~\simeq~ {2 m_q \over \fpi^2(T) } \ts
| \langle \bar{q} q \rangle_T |,
\eeq
where (again, for two massless flavors)~\cite{temp}
\beq
\fpi^2(T) ~=~  \fpi^2(0)
\Biggl[ 1 - {T^2 \over 6 \fpi^2(0)} + {\cal O}(T^4) \Biggr].
\eeq
Note that at leading order the thermal pion mass actually increases
slightly with temperature.
The temperature dependence of the baryon masses is a harder problem,
but in the na{\"\i}ve quark model we
might expect them to behave roughly as
\beq
\label{xqm}
m_B (T) ~\simeq~ 3  ~ | \langle \bar{q} q \rangle_T | ,
\eeq
where the constituent quark mass is simply due to the condensate.
This estimate at least incorparates the fact that
the baryons must become nearly massless at the chiral
phase boundary, since chiral symmetry prevents them from obtaining
a mass.   
Equations~(\ref{DT}) and~(\ref{xqm}) suggest that the relative abundance of
baryons compared to pions will increase at high temperature.
While we don't necessarily believe that (\ref{xqm}) is very accurate,
the point is that the thermal pion and baryon masses probably do not
depend on the quark condensate (and hence the temperature) in the same way.
Thus, it is probably inconsistent to imagine fitting the properties of an
interacting hadron gas at $T \sim m_\pi$ using vacuum hadron masses.
Yet, essentially all recent heavy ion data\footnote{In fact, from this
point of view any model (such as a parton cascade model) which
reproduces free particle thermal multiplicites at temperatures of order
150 MeV probably lacks some important dynamics associated with the phase
transition. We might classify it as just another efficient populator of
phase space.} agrees well with
multiplicities generated by free thermal models, with temperatures of
roughly $T \sim 50 - 170$~MeV.


\begin{figure}[t!]
\includegraphics{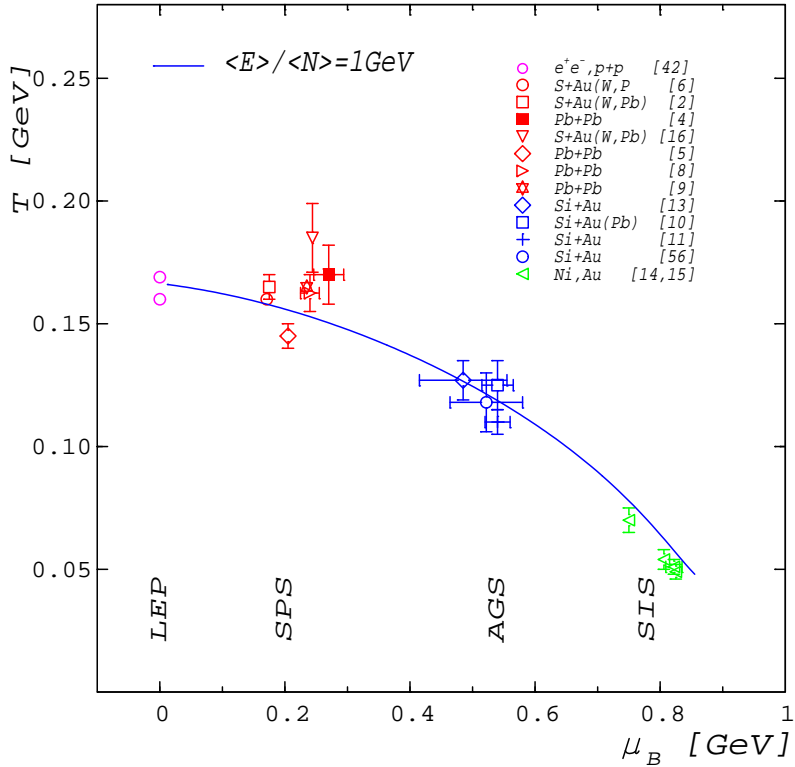}
\vspace{3.8in}
\caption{
Temperatures and chemical potentials obtained by
Cleymans and Redlich~\protect\cite{BH8} from thermal
multiplicity fits of LEP, CERN/SPS, BNL/AGS and
GSI/SIS data.}
\label{phasediag}
\end{figure}

In Fig.~\ref{phasediag} we reproduce a plot from the paper of
Cleymans and Redlich~\cite{BH8}, where the fitted temperatures and
chemical potentials resulting from LEP, CERN/SPS, BNL/AGS and
GSI/SIS data are displayed. One interpretation of these results
(which we do not subscribe to) is that a large region of the QCD
phase diagram in the temperature-density plane has already been
explored! There is no question that the quality of the free
thermal fits is quite good. However, this in itself suggests that
an interacting thermal region has yet to be produced in these
experiments. Rather, it is quite possible that the collisions
simply serve as a mechanism for populating phase space, without
ever evolving through configurations in real thermal and chemical
equilibrium ({\it i.e.} actual points on the phase diagram in
Fig.~\ref{phasediag}). If the system {\it had} passed through real
equilibrium, we suggest that the observed final state
multiplicities could deviate significantly from those which can be
generated by free thermal models. Thus the failure of such models
to fit the data could be a signal for real equilibrium. This idea
is explored in the context of phenomenologically motivated models
in \cite{PH}. For more recent work related to this paper, which
originally appeared as preprint nucl-th/0001044 in 2000, see
\cite{recent}.

\bigskip

\section*{Acknowledgements}
\noindent The authors would like to thank S.~Das~Gupta, C.~Gale,
R.~Hwa, C.S.~Lam, H.~Minakata, B.~Mueller, R.~Pisarski and
D.~Rischke for useful discussions and comments. SH is particularly
grateful to K.S. Lee for making him aware of the previous work of
C.N. Yang and collaborators, and for re-stimulating his interest
in this area. SH is supported under DOE contract
DE-FG06-85ER40224. JH and GM are supported in part by the Natural
Sciences and Engineering Research Council of Canada and the Fonds
pour la Formation de Chercheurs et l'Aide \`a la Recherche of
Qu\'ebec.


\bigskip



\vskip 1 in
\baselineskip=1.6pt
\begin{thebibliography}{99}
%
%
\def\np#1#2#3{  {Nucl. Phys. #1} (19#3) #2}
\def\pl#1#2#3{  {Phys. Lett. #1} (19#3) #2}
\def\pr#1#2#3{   {Phys. Rev. #1} (19#3) #2}
\def\prep#1#2#3{ {Phys. Rep. #1} (19#3) #2)}
\def\prl#1#2#3{ {Phys. Rev. Lett. #1} (19#3) #2}
%
\bibitem{old} E. Fermi, Prog. Theor. Phys. {\bf 5}, 570
(1950); L.D. Landau, Izv.Akad.Nauk Ser. Fiz. {\bf 17}, 51 (1953);
R. Hagedorn, Nuovo Cimento Suppl. {\bf 3}, 147 (1965).

\bibitem{BH1}\label{BH1}
For a review, see, {\it e.g.},
U.~Heinz, J. Phys. {\bf G25}, 263 (1999);
talk presented at the 14th International Conference on
Ultrarelativistic Nucleus-Nucleus Collisions,
Torino, Italy, May 1999, nucl-th/9907060.


\bibitem{BH3}
F.~Becattini, Z. Phys. {\bf C69}, 485 (1996).

\bibitem{BH4}
F.~Becattini, Firenze Preprint DFF 263/12/1996,
hep-ph/9701275.

\bibitem{BH5}
F.~Becattini and U.~Heinz, Z. Phys. {\bf C76}, 269 (1997).

\bibitem{BH2}
J.~Sollfrank, M.~Gazdzicki, U.~Heinz and J.~Rafelski,
Z. Phys. {\bf C61}, 659 (1994).

\bibitem{BH11}
P.~Braun-Munzinger, J.~Stachel, J.P.~Wessels and N.~Xu,
Phys. Lett. {\bf B365}, 1 (1996).

\bibitem{BH10}
P.~Braun-Munzinger and J.~Stachel,
Nucl. Phys. {\bf A606}, 320 (1996).

\bibitem{BH6}
F.~Becattini, M.~Ga\'zdzicki and J.~Sollfrank,
Nucl. Phys. {\bf A638}, 403 (1998);
Eur. Phys. J. {\bf C5}, 143 (1998).

\bibitem{BH7}
J.~Sollfrank, Eur. Phys.~J. {\bf C9}, 159 (1999).

\bibitem{BH9}
P.~Braun-Munzinger, I.~Heppe and J.~Stachel,
Phys. Lett. {\bf B465}, 15 (1999).

\bibitem{BH8}\label{BH8}
J.~Cleymans and K.~Redlich, Phys. Rev. {\bf C60}, 054908 (1999).


\bibitem{Yang} T.T. Chou, C.N. Yang and E. Yen, Phys.Rev.Lett.
{\bf54}, 510 (1985);T.T. Chou and C.N. Yang, Phys.Rev.Lett.
{\bf55}, 1359 (1985).



\bibitem{QED}
S.J.~Parke and T.R.~Taylor,
Phys. Rev. Lett. {\bf 56}, 2459 (1986);
G.~Mahlon and \hbox{T.--M.}~Yan, Phys. Rev. {\bf D47}, 1776 (1993).

\bibitem{lattice}
Y.~Nakahara, M.~Asakawa, and T.~Hatsuda,
talk given at the 17th International Symposium on Lattice Field
Theory (LATTICE 99), Pisa, Italy, hep-lat/9909137.

\bibitem{temp}
J.~Gasser and H.~Leutwyler, Phys. Lett. {\bf B184}, 83 (1987);
P.~Gerber and H.~Leutwyler, Nucl. Phys. {\bf B321}, 387 (1989).

\bibitem{PH}
S. Pratt and K. Haglin, Phys. Rev. {\bf C59}, 3304 (1999).

\bibitem{recent} D. Rischke, NPA {\bf 698}, 153C (2002); V. Koch
nucl-th/0210070; U. Heinz, nucl-th/0212004; J. Rafelski,
nucl-th/0212091.



\end{thebibliography}
\end{document}